 \documentclass[final,5p,times,twocolumn]{elsarticle}

 \usepackage{graphics}
 \usepackage{graphicx}
\usepackage{amssymb}
\journal{Journal of Quantitative Spectroscopy and Radiative Transfer}

\begin{document}

\begin{frontmatter}



\title{Spectroscopic parameters of phosphine, PH$_3$, in its ground vibrational state}


\author[Koeln]{Holger S.P. M\"uller\corref{cor}}
\ead{hspm@ph1.uni-koeln.de}
\cortext[cor]{Corresponding author.}

\address[Koeln]{I.~Physikalisches Institut, Universit{\"a}t zu K{\"o}ln, 
   Z{\"u}lpicher Str. 77, 50937 K{\"o}ln, Germany}

%%%%%%%%%%%%%%%%%%%%%%%%%%%%%%%%%%%%%%%%%%%%%%%%%%%%%%%%%%%%%%%%%%%%%%%%%%%%%%%%%%%%%
%%%%%%%%%%%%%%%%%%%%%%%%%%%%%%%%%%%%%%%%%%%%%%%%%%%%%%%%%%%%%%%%%%%%%%%%%%%%%%%%%%%%%
%%%%%%%%%%%%%%%%%%%%%%%%%%%%%%%%%%%%%%%%%%%%%%%%%%%%%%%%%%%%%%%%%%%%%%%%%%%%%%%%%%%%%
\begin{abstract}

The ground state rotational spectrum of PH$_3$ has been reanalyzed taking into account 
recently published very accurate data from sub-Doppler and conventional absorption 
spectroscopy measurements as well as previous data from the radio-frequency to the 
far-infrared regions. These data include $\Delta J = \Delta K = 0$ transitions between 
$A_1$ and $A_2$ levels, $\Delta J = 0$, $\Delta K = 3$ transitions as well as regular 
$\Delta J = 1$, $\Delta K = 0$ rotational transitions. Hyperfine splitting caused by 
the $^{31}$P and $^1$H nuclei has been considered, and the treatment of the $A_1$/$A_2$ 
splitting has been discussed briefly. Improved spectroscopic parameters have been 
obtained. Interestingly, the most pronounced effects occured for the hyperfine parameters. 

\end{abstract}


\end{frontmatter}

%% \linenumbers %%% tried, but did not work on my latex distribution

%%%%%%%%%%%%%%%%%%%%%%%%%%%%%%%%%%%%%%%%%%%%%%%%%%%%%%%%%%%%%%%%%%%%%%%%%%%%%%%%%%%%%
%%%%%  main text  %%%%%%%%%%%%%%%%%%%%%%%%%%%%%%%%%%%%%%%%%%%%%%%%%%%%%%%%%%%%%%%%%%%
%%%%%%%%%%%%%%%%%%%%%%%%%%%%%%%%%%%%%%%%%%%%%%%%%%%%%%%%%%%%%%%%%%%%%%%%%%%%%%%%%%%%%

%%%%%%%%%%%%%%%%%%%%%%%%%%%%%%%%%%%%%%%%%%%%%%%%%%%%%%%%%%%%%%%%%%%%%%%%%%%%%%%%%%%%%
%%%%%  Introduction  %%%%%%%%%%%%%%%%%%%%%%%%%%%%%%%%%%%%%%%%%%%%%%%%%%%%%%%%%%%%%%%%
%%%%%%%%%%%%%%%%%%%%%%%%%%%%%%%%%%%%%%%%%%%%%%%%%%%%%%%%%%%%%%%%%%%%%%%%%%%%%%%%%%%%%

\section{Introduction}
\label{introduction}

Phosphine, PH$_3$, was detected as a trace constituent of the atmospheres of the 
giant planets Jupiter~\cite{PH3-Jupiter_1976} and Saturn~\cite{PH3-Saturn_1977}, 
respectively by infrared spectroscopy. Observations of the lowest $J = 1 - 0$ 
rotational transition near 267~GHz toward Jupiter~\cite{Jupiter_J=1-0_1984} and 
Saturn~\cite{Saturn_J=1-0_1994} were also reported. Using a far- and mid-infrared 
(FIR and MIR) spectrometer on board of the \textit{Cassini} satellite, 
it was even possible to study the meridional PH$_3$ distribution in Saturn's upper 
troposphere~\cite{Saturn_Cassini_2007}. Higher resolution FIR observations of Saturn's 
atmosphere were carried out very recently with the \textit{Herschel} satellite, 
and the vertical distribution of PH$_3$ was derived~\cite{Saturn_Herschel_2012}. 
Phosphine is also a likely species to occur in the atmospheres of giant extrasolar 
planets or of brown dwarfs~\cite{opacities4brown-dwarfs_etc_2007}. 
Moreover, it has been tentatively detected in the circumstellar shells of carbon-rich 
late-type stars CW~Leonis~\cite{tent_det_IRC1_2008,tent_det_IRC2_2008}, also known as 
IRC~+10216, and V1610~Cygni, also known as CRL~2688, by radio-astronomical observations 
of the $J = 1 - 0$ transition.

Small molecules, and even more so light hydride species, such as PH$_3$, are also of 
great interest for basic science. In the early years, an important questions was if 
PH$_3$ would display measureable inversion splitting as its lighter homolog ammonia, 
NH$_3$, does~\cite{NH3-inversion_1932a,NH3-inversion_1932b}. 

Early reports on the ground state rotational spectrum of PH$_3$ were published by the 
Gordy group~\cite{PH3_rot_1954,PH3_rot_1969}. Burrus~\cite{PH3_Stark_1958} determined 
the dipole moment from Stark measurements. Davies et al.~\cite{PH3_RF_1971} studied 
transitions between $A_1$ and $A_2$ levels with $\Delta J = \Delta K = 0$ at 
radio-frequencies (RF). They were unable to resolve any tunneling splitting, but obtained 
hyperfine structure parameters caused by the $^{31}$P and $^1$H nuclei and an improved 
value for the dipole moment. Chu and Oka recorded $\Delta K = 3$ transitions with 
$\Delta J = 0$ and $K = \pm1 - \mp2$ and estimated an effective value for the centrifugal 
distortion induced dipole moment from intensity measurements~\cite{PH3_rot_DelK=3_1974}. 
Later, corresponding transitions were reported for $K = 0 - 3$~\cite{PH3_rot_DelK=3_1977}, 
$K = 2 - 5$~\cite{PH3_rot_DelK=3_1981a}, and $K = 3 - 6$~\cite{PH3_rot_DelK=3_1981b}. 
Regular rotational transitions with $\Delta J = 1$ and $\Delta K = 0$ in the millimeter 
and submillimeter regions~\cite{PH3_rot_DelK=3_1981a,PH3_rot_DelK=3_1981b} and in the 
far-infrared (FIR) region~\cite{PH3_rot_FIR_1988} were also studied. 

Recently, Cazzoli and Puzzarini~\cite{PH3_rot_Lamb-Dip_2006} obtained very accurate 
transition frequencies for the $J = 1 - 0$ and $2 - 1$ transitions employing 
sub-Doppler (Lamb-dip) spectroscopy, which resolved $^{31}$P and $^1$H hyperfine 
structure, and for the $J = 3 - 2$ transitions using conventional absorption
spectroscopy. As this study employed in their fit very limited data beyond their own, 
a combined fit of these data together with previous results has been carried out here 
to obtain an improved set of spectroscopic parameters which should be useful not only 
for spectroscopic observations of phosphine in planetary atmospheres, brown dwarfs, 
or other regions in space, but also for future analyses of rovibrational spectra or 
as benchmarks for quantum-chemical calculations.

%%%%%%%%%%%%%%%%%%%%%%%%%%%%%%%%%%%%%%%%%%%%%%%%%%%%%%%%%%%%%%%%%%%%%%%%%%%%%%%%%%%%%
%%%%%  Spectral analysis  %%%%%%%%%%%%%%%%%%%%%%%%%%%%%%%%%%%%%%%%%%%%%%%%%%%%%%%%%%%
%%%%%%%%%%%%%%%%%%%%%%%%%%%%%%%%%%%%%%%%%%%%%%%%%%%%%%%%%%%%%%%%%%%%%%%%%%%%%%%%%%%%%

%%%%%%%%%%%%%%%%%%%%%%%%%%%%%%%%%%%%%%%%%%%%%%%%%%%%%%%%%%%%%%%%%%%%%%%%%%%%%%%%%%%%%
%%%%%  Table 1  %%%%%%%%%%%%%%%%%%%%%%%%%%%%%%%%%%%%%%%%%%%%%%%%%%%%%%%%%%%%%%%%%%%%%
%%%%%%%%%%%%%%%%%%%%%%%%%%%%%%%%%%%%%%%%%%%%%%%%%%%%%%%%%%%%%%%%%%%%%%%%%%%%%%%%%%%%%

\begin{table*}
  \caption{Spectroscopic parameters$^a$ (MHz) of phosphine, PH$_3$, from the present fits in 
           comparison to selected previous fits.}
  \label{parameters}
{\footnotesize
  \begin{tabular}{lr@{}lr@{}lr@{}lr@{}lr@{}lr@{}l}
  \hline
    & \multicolumn{2}{l}{Present, preferred$^b$} & \multicolumn{2}{l}{Present, alternative$^c$} & \multicolumn{2}{l}{Fusina and Carlotti$^d$} & 
\multicolumn{2}{l}{Cazzoli and Puzzarini$^e$} & \multicolumn{2}{l}{Davies et al.$^f$} \\
  \hline
$B$                            & 133\,480&.128\,21~(15)    & 133\,480&.127\,94~(15)    & 133\,480&.122\,80~(177)   & 133\,480&.128\,989~(95) &         &$-$      \\
$C$                            & 117\,489&.430\,4~(73)$^g$ & 117\,489&.427\,0~(73)$^g$ & 117\,489&.425\,6~(73)     & 117\,489&.436$^{g,h}$   &         &$-$      \\
$D_J$                          &        3&.936\,391~(22)   &        3&.936\,210~(22)   &        3&.936\,059~(105)  &        3&.936\,901~(36) &         &$-$      \\
$D_{JK}$                       &     $-$5&.168\,535~(78)   &     $-$5&.167\,110~(78)   &     $-$5&.167\,280~(86)   &     $-$5&.171\,02~(44)  &         &$-$      \\
$D_K$                          &        4&.234\,45~(54)    &        4&.232\,60~(54)    &        4&.233\,82~(51)    &         &$-$            &         &$-$      \\
$H_J \times 10^6$              &      348&.906~(142)       &      326&.315~(142)       &      325&.74~(39)         &      415&.7~(39)        &         &$-$      \\
$H_{JK} \times 10^6$           &   $-$868&.02~(50)         &   $-$664&.97~(50)         &   $-$665&.33~(73)         &$-$1\,237&.~(85)         &         &$-$      \\
$H_{KJ} \times 10^6$           &      801&.49~(148)        &      463&.43~(149)        &      464&.10~(276)        &   1\,340&.~(270)        &         &$-$      \\
$H_K \times 10^6$              &    $-$31&.4~(111)         &      114&.8~(111)         &      152&.7~(103)         &         &$-$            &         &$-$      \\
$h_3 \times 10^6$              &       11&.3$^b$           &         &$-^c$            &         &$-$              &         &$-$            &         &$-$      \\
$L_J \times 10^9$              &    $-$37&.667~(218)       &    $-$37&.672~(218)       &    $-$36&.96~(43)         &         &$-$            &         &$-$      \\
$L_{JJK} \times 10^9$          &      113&.71~(105)        &      149&.91~(106)        &         &$-$              &         &$-$            &         &$-$      \\
$L_{JK} \times 10^9$           &   $-$222&.0~(33)          &   $-$408&.7~(33)          &      349&.0~(39)          &         &$-$            &         &$-$      \\
$L_{KKJ} \times 10^9$          &      247&.9~(50)          &      519&.5~(50)          &   $-$547&.7~(93)          &         &$-$            &         &$-$      \\
$\epsilon \times 10^3$         &      714&.508\,2~(42)$^i$ &      831&.652\,2~(36)$^i$ &      829&.254~(56)        &         &$-$            &         &$-$      \\
$\epsilon _J \times 10^6$      &   $-$168&.896~(72)$^i$    &   $-$180&.210~(62)$^i$    &   $-$179&.611~(204)       &         &$-$            &         &$-$      \\
$\epsilon _K \times 10^6$      &         &$-$              &         &$-$              &      265&.9~(61)          &         &$-$            &         &$-$      \\
$\epsilon _{JJ} \times 10^9$   &       37&.25~(29)$^i$     &       34&.60~(24)$^i$     &       34&.97~(199)        &         &$-$            &         &$-$      \\
$\epsilon _{JK} \times 10^9$   &         &$-$              &         &$-$              &    $-$71&.5~(102)         &         &$-$            &         &$-$      \\
$\epsilon _{KK} \times 10^9$   &         &$-$              &         &$-$              &       87&.2~(82)          &         &$-$            &         &$-$      \\
$C_N(P) \times 10^3$           &      114&.897~(15)        &      114&.897~(15)        &         &$-$              &      115&.35~(12)       &      114&.90~(13) \\
$C_K(P) \times 10^3$           &      116&.396~(35)$^g$    &      116&.396~(35)$^g$    &         &$-$              &      115&.0~(14)$^g$    &      116&.38~(32) \\
$C_N(H) \times 10^3$           &     $-$7&.998~(9)         &     $-$7&.998~(9)         &         &$-$              &     $-$7&.57~(13)       &     $-$8&.01~(8)  \\
$C_K(H) \times 10^3$           &     $-$7&.704~(22)$^g$    &     $-$7&.704~(22)$^g$    &         &$-$              &     $-$7&.69$^{g,i}$    &     $-$7&.69~(19) \\
$-3D_1$(P$-$H$) \times 10^3$   &        3&.01~(35)         &        3&.01~(35)         &         &$-$              &        3&.05~(62)       &         &$k$      \\
$-0.5D_2$(P$-$H$) \times 10^3$ &     $-$9&.15$^l$          &     $-$9&.15$^l$          &         &$-$              &     $-$9&.15$^l$        &         &$k$      \\
$1.5D_3$(H$-$H$) \times 10^3$  &      20&.62~(13)          &       20&.62~(13)         &         &$-$              &       25&.3~(19)        &         &$k$      \\
    \hline \hline
  \end{tabular}\\[2pt]
}
$^a$\footnotesize{Numbers in parentheses are one 
    standard deviation in units of the least significant figures.}\\
$^b$\footnotesize{The parameter $h_3$ was kept fixed to the value calculated in Ref.~\cite{PH3_analysis_1985}.}\\
$^c$\footnotesize{The parameter $h_3$ was constrained to zero.}\\
$^d$\footnotesize{Ref.~\cite{PH3_rot_FIR_1988}, preferred fit with $h_3$ etc. kept fixed to zero.}\\
$^e$\footnotesize{Ref.~\cite{PH3_rot_Lamb-Dip_2006}.}\\
$^f$\footnotesize{Ref.~\cite{PH3_RF_1971}.}\\
$^g$\footnotesize{The parameters $C$ was determined from $C - B$ and $B$, taking correlation into account. 
     Analogously, $C_K$ was calculated from $C_K - C_N$ and $C_N$.}\\
$^h$\footnotesize{$C$ was kept fixed to the value in Ref.~\cite{PH3_analysis_1985}.}\\
$^i$\footnotesize{The parameters determined in the fit are $\sqrt{2} \epsilon$ etc.}\\
$^j$\footnotesize{$C_K(H)$ was kept fixed to the value in Ref.~\cite{PH3_RF_1971}.}\\
$^k$\footnotesize{The tensorial spin-spin coupling parameters were kept fixed to values calculated 
     from structural parameters, but were not given specifically.}\\
$^l$\footnotesize{$D_2$ was kept fixed to the value calculated in Ref.~\cite{PH3_rot_Lamb-Dip_2006}.}\\
\end{table*}

%%%%%%%%%%%%%%%%%%%%%%%%%%%%%%%%%%%%%%%%%%%%%%%%%%%%%%%%%%%%%%%%%%%%%%%%%%%%%%%%%%%%%
%%%%%%%%%%%%%%%%%%%%%%%%%%%%%%%%%%%%%%%%%%%%%%%%%%%%%%%%%%%%%%%%%%%%%%%%%%%%%%%%%%%%%
%%%%%%%%%%%%%%%%%%%%%%%%%%%%%%%%%%%%%%%%%%%%%%%%%%%%%%%%%%%%%%%%%%%%%%%%%%%%%%%%%%%%%

\section{Spectral analysis and discussion}
\label{a and d}

Phosphine is an oblate rotor with a permanent dipole moment of 
0.57395~(3)~D~\cite{PH3_RF_1971}. It is moderately close to the spherical rotor limit 
with $\gamma = (C_0 - B_0)/A_0 = -0.1361$ being small in magnitude, but not particularly so.

The rotational energies and transitions of a strongly oblate rotor in its ground 
vibrational state and without $A_1$/$A_2$ splitting require for their modeling only 
$C - B$ and $B$ along with distortion parameters, $D_J$, $D_{JK}$, and $D_K$ to lowest 
order. In a strongly prolate rotor, $C$ is replaced with $A$. The purely axial parameters 
$C - B$, $D_K$, etc. can usually not be determined by rotational spectroscopy. 
Their determination requires, e.g., the observation of centrifugal distortion induced 
$\Delta K = 3$ transitions. 
The $A_1$/$A_2$ splitting is modeled to lowest order by $h_3 (J_+^6 + J_-^6)$ with 
$J_{\pm} = J_x \pm iJ_y$. $J$ and $K$ distortion corrections and possibly even 
higher order terms may be required in the fit. Early treatments of the rotational 
spectrum of PH$_3$ used parameters as described above to fit the rotational transitions. 
However, it was recognized early on that the parameter derived for $h_3$ was too large 
by about a factor of 4~\cite{PH3_RF_1971}. Moreover, it was found later, that the 
Hamiltonian converged rather slowly~\cite{PH3_rot_DelK=3_1981b}.

Tarrago and Dang Nhu~\cite{PH3_analysis_1985} discussed the formulation of the Hamiltonian 
for symmetric top molecules approaching the spheric rotor limit employing the examples 
OPF$_3$ and PH$_3$. They pointed out that for such molecules not only the $\Delta K = 6$ 
terms derived from $h_3$, also called $f_6$ in their work, affect the $A_1$/$A_2$ splitting, 
but also terms with $\Delta K = 3$, whose lowest order term is $\epsilon$, 
which is also called $q_3$. The parameter $\epsilon$ is the coefficient of 
$[J_+^3 + J_-^3,J_z]_+$, were $[A,B]_+ = AB + BA$ is the anticommutator of $A$ and $B$. 
They presented one fit with parameters derived from $h_3$, and one fit with parameters 
derived from $\epsilon$ and $h_3$. The latter fit was slightly better, but yielded 
a value for $h_3$ only slightly smaller than in the fit without $\epsilon$, meaning that 
their value was still too lage by almost a factor of 4. Moreover, their value for 
$\epsilon$ was too small by about a factor of 3. Interestingly, they did not attempt 
any fits without parameters derived from $h_3$.

Fusina and Carlotti recorded the FIR spectrum of PH$_3$ and made assignments for 
the ground vibrational state between 44 and 202~cm$^{-1}$ with $J''$ between 4 and 22 
and $K$ between 19 and 0~\cite{PH3_rot_FIR_1988}. They used in their fits their own data 
together with previous data from 
Refs.~\cite{PH3_rot_1969,PH3_RF_1971,PH3_rot_DelK=3_1974,PH3_rot_DelK=3_1977,PH3_rot_DelK=3_1981a,PH3_rot_DelK=3_1981b}. 
Three different fits were presented which included parameters derived from both $\epsilon$ 
and $h_3$ as well as parameters derived either from $\epsilon$ or $h_3$. The parameters 
$\epsilon$ and $h_3$ were completely correlated if both were used. The parameters 
$h_3$ and $h_{3K}$ were completely correlated in the fit without parameters derived 
from $\epsilon$. In contrast, the fit with only parameters derived from $\epsilon$ did 
not produce severe correlations, yielded the best standard deviation, but employed two 
more parameters than the other two fits. This fit was their preferred one.

The fits carried out in the course of the present investigation employed Herb Pickett's 
{\scriptsize SPFIT} and {\scriptsize SPCAT} programs~\cite{Pickett_1991}. They started 
from the preferred parameters from Fusina and Carlotti~\cite{PH3_rot_FIR_1988} and 
the hyperfine parameters employed by Cazzoli and Puzzarini~\cite{PH3_rot_Lamb-Dip_2006} 
using transition frequencies from 
Refs.~\cite{PH3_RF_1971,PH3_rot_DelK=3_1974,PH3_rot_DelK=3_1977,PH3_rot_DelK=3_1981a,PH3_rot_DelK=3_1981b,PH3_rot_FIR_1988,PH3_rot_Lamb-Dip_2006}
with uncertainties as reported. The RF transitions~\cite{PH3_RF_1971} were reportedly 
accurate to between 0.1 and 0.2~kHz. Initially, 0.2~kHz were used in the fits because 
the published deviations from the measured frequencies exceeded 0.6~kHz in two cases. 
However, the transitions were reproduced in the present fits to well within 0.1~kHz on 
average with no severe outliers. Therefore, 0.1~kHz uncertainties were employed 
for these data in the final fits. The uncertainties of Ref.~\cite{PH3_rot_DelK=3_1981a} 
were used as reported in Ref.~\cite{PH3_rot_DelK=3_1981b}. Including $L_{JJK}$ in the fit, 
which was contrained to zero in Ref.~\cite{PH3_rot_FIR_1988}, yielded a reasonable value, 
and suggested the parameters $\epsilon _K$, $\epsilon _{JK}$, and $\epsilon _{KK}$ to be 
indeterminate. The latter three parameters could be omitted from the fit without 
significant deterioration of the fit. The resulting parameters, including those for the 
hyperfine structure, are given in Table~\ref{parameters} under heading ''Present, alternative''. 

Hyperfine structure caused by the $^{31}$P and $^1$H nuclei was resolved in the RF 
spectra~\cite{PH3_RF_1971} and in the Lamb-dip measurements from Ref.~\cite{PH3_rot_Lamb-Dip_2006}. 
The hyperfine Hamiltonian was described in sufficient detail e.g. in the latter work. 
As mentioned in that work, the Lamb-dip measurements were rather sensitive to the 
perpendicular components $C_N$ of the P and H nuclear spin-rotation tensors, which are 
also known as $C_{\perp}$, and to the axial components $D_1$(P$-$H) and $D_3$(H$-$H) of 
the spin-spin coupling tensors, which are also known as $d_{\parallel}$(P$-$H) and 
$d_{\parallel}$(H$-$H), respectively. They were not as sensitive to the axial components 
$C_K$ of the nuclear spin-rotation tensor, which are also known as $C_{\parallel}$, and 
to the perpendicular component $D_2$(P$-$H) of the P$-$H spin-spin coupling tensor, 
also known as $d_{\perp}$(P$-$H). In contrast, the RF data were sensitive to $C_N$ and 
$C_K$ of both nuclei, but not as sensitive to the spin-spin coupling tensors. 
In the present fit, $D_2$ could still not be determined significantly and with confidence. 
However, its omission increased the standard deviation of the fit somewhat. Therefore, 
it was retained in the fit kept fixed to the calculated value from 
Ref.~\cite{PH3_rot_Lamb-Dip_2006}. Other parameters, such as $C_-$(H), the off-diagonal 
H spin-rotation parameter, the scalar spin-spin coupling parameters, or distortion 
corrections to the P or H nuclear spin-rotation parameters were not determined with 
significance, and their inclusion had a negligible effect on the quality of the fit. 
Therefore, they were omitted from the final fits.

As can be seen in Table~\ref{parameters}, the present hyperfine parameters are in good to 
very good agreement with previously determined ones~\cite{PH3_RF_1971,PH3_rot_Lamb-Dip_2006}, 
but have much smaller uncertainties in the present fits. The values of the spin-spin 
coupling parameters $-3D_1$ and $1.5D_3$ agree very well with the calculated ones of 
3.18~kHz and 20.7~kHz, respectively. In fact, keeping the spin-spin coupling parameters 
fixed to the calculated values, does not deteriorate the quality of the fit significantly. 
In addition, the values of the other parameters and, somewhat surprisingly, even their 
uncertainties are affected negligibly. 

The remaining spectroscopic parameters agree for the most part well with the preferred 
ones from Fusina and Carlotti~\cite{PH3_rot_FIR_1988} as far as they were determined 
in both fits. Exceptions are the two octic distortion parameters $L_{JK}$ and $L_{KKJ}$, 
which is a result of the omission of $L_{JJK}$ by Fusina and Carlotti and of the omission 
of some higher order distortion corrections to $\epsilon$ in the present fits. Notably, 
the values of $\epsilon$ agree quite well, about 832~kHz versus 829~kHz. Both values, 
however, are slightly larger than 723~kHz calculated by Tarrago and Dang 
Nhu~\cite{PH3_analysis_1985} and 773~kHz calculated by Fusina and 
Carlotti~\cite{PH3_rot_FIR_1988}. The parameter $\epsilon$ and its distortion corrections 
cause $\Delta K = 3$ transitions, among others, to have non-zero intensities very similar 
to those calculated without these parameters, but with the effective centrifugal distortion 
correction to the permanent dipole moment mentioned in Ref.~\cite{PH3_rot_DelK=3_1974}. 
Therefore, another fit has been carried out with $h_3$ kept fixed at the value 
of 11.3~Hz calculated in Ref.~\cite{PH3_analysis_1985}. The $\epsilon$ value of 
714.5~kHz resulting from this fit is very close to the one calculated by Tarrago and 
Dang Nhu~\cite{PH3_analysis_1985}, but somewhat smaller than the one from Fusina and 
Carlotti~\cite{PH3_rot_FIR_1988}. 
All other parameters are also affected in their magnitudes. This fit is considered to 
be the preferred one. The effects on the intensities of the $\Delta K = 3$ transitions 
correspond closely to the square of the ratios of the $\epsilon$ values, the intensities 
of all other transitions change only marginally between the two data sets.

The quality of the preferred fit and of the alternative fit are identical. The experimental 
transition frequencies have been reproduced within the reported uncertainties on average, 
the rms error is 0.93, slightly better than the ideal 1.0. There is some scatter among 
the rms errors of individual data sets: it is 0.77 for the RF data~\cite{PH3_RF_1971} 
and 0.64~\cite{PH3_rot_DelK=3_1974}, 0.87~\cite{PH3_rot_DelK=3_1977}, and 
0.94~\cite{PH3_rot_DelK=3_1981a,PH3_rot_DelK=3_1981b} for the older millimeter and 
submillimeter data. The FIR data from Fusina and Carlotti~\cite{PH3_rot_FIR_1988} 
were given uncertainties of 0.0001~cm$^{-1}$ throughout, as suggested by the authors, 
even though the residuals increased somewhat for the transitions with the highest two 
$J$ values. Two transitions omitted already by Fusina and Carlotti were also omitted here. 
In addition, the frequency of the transition with $J'' = 16$ and $K = 9$ was omitted as 
its residual exceeded $4\sigma$. The rms error of the FIR transitions was 0.94; 
the frequencies of 4 transitions, all at higher $J$ values, had residuals larger than 
$3\sigma$, which is reasonably close to the statistical expectations in the case of 230 
different transition frequencies. An rms error of 1.37 was obtained for the data from 
Ref.~\cite{PH3_rot_Lamb-Dip_2006}. Even though this suggests slightly optimistic 
uncertainty estimates, no adjustments were made, as one frequency, 533815.1259~(10)~MHz, 
contributed particularly to the rms error. Omission of this line reduced the rms error 
of this data set to slightly below 1.1. In addition, the omission of this line from 
the overall line list affected all parameters within experimental uncertainties. 
Therefore, this line was retained in the final fit. 

The $J$-dependent parameters, in particular the ones dependent purely on $J$ have much 
smaller uncertainties in the present fits compared to the preferred fit of Fusina 
and Carlotti~\cite{PH3_rot_FIR_1988}. This is hardly surprising given the nature of 
the new data from Ref.~\cite{PH3_rot_Lamb-Dip_2006}. The uncertainties of $D_K$ and 
$H_K$ here are actually slightly larger than from the FIR fit. This may be caused by 
differences in the correlation of the parameters or by differences in the weighting 
scheme.

Predictions of the rotational spectrum were made for the Cologne Database for Molecular 
Spectroscopy, CDMS\footnote{http://www.astro.uni-koeln.de/cdms/}~\cite{CDMS_1,CDMS_2}, 
using the parameters of the preferred fit. The upper quantum numbers were 
$J = 40$ and $K = 38$, ensuring convergence of the rotational part of the 
partition function to $10^{-4}$ up to 500~K. The intensity cut-offs were chosen 
to include all experimental $\Delta J = \Delta K = 0$ transitions between 
$A_1$ and $A_2$ levels on one hand and to limit the number of transitions 
with uncertainties equal to or larger than 1~GHz, as {\scriptsize SPCAT} displays 
such uncertainties as 999.9999~MHz or 0.0334~cm$^{-1}$. This leads to transitions 
with upper $J$ and $K$ of 34 and 31, respectively. The $K = 0$ transition with 
$J = 8 - 7$ is the most intense transition at 300~K. At 500~K and 1000~K, the 
corresponding transition with $J = 11 - 10$ and $J = 14 - 14$, respectively, is the 
most intense one. With phosphine being only a minor constituent in the atmospheres 
of Jupiter and Saturn, the current predictions should therefore contain sufficient 
ground state rotational transitions not only at 300~K, but also, after appropriate 
conversion of the intensities, for 500~K and likely for 1000~K. However, the neglect 
of rotational transitions in excited vibrational states may cause non-negligible errors 
already around 300~K, as can be seen below.

Intensities of the transitions with $\Delta K = 3$ should be viewed with some caution 
as indicated above. Predictions of transition frequencies with $\Delta K = 0$ should be 
reliable as long as the predicted uncertainties do not exceed 5~MHz. Predictions of 
transition frequencies with $\Delta K \not= 0$ should be viewed with caution if the 
predicted uncertainties are larger than 0.5~MHz. In these cases, the actual frequencies 
may differ from the predictions by more than five times the predicted uncertainties.

The partition function values given in the documentation of the CDMS 
entry\footnote{http://www.astro.uni-koeln.de/cgi-bin/cdmsinfo?file=e034501.cat} do not 
include vibrational contributions; they are about 2.4\,\% at 300~K and amount to about 
18.5\,\% at 500~K. Classically estimated vibrational contributions are intended to 
be provided through a link in the documentation file. 
It should be mentioned that the HITRAN entry takes into account vibrational contributions 
to the partition function. Moreover, the intensities at 300~K have been converted to 
those at 296~K, which is the default temperature in HITRAN.

Pickett's {\scriptsize SPFIT} and {\scriptsize SPCAT} programs~\cite{Pickett_1991} 
use the total parity. This parity is therefore also used in the {\scriptsize ASCII} 
table version of the PH$_3$ entry in the CDMS. It differs both from the $A_1$/$A_2$ 
parity as well as from the $A_+$/$A_-$ parity. $A$ levels with $-$ parity belong to 
$A_2$ for even $K$ and to $A_1$ for odd $K$. $A$ levels with $-$ parity belong to 
$A_-$ for $J + K$ even and to $A_+$ for $J + K$ odd. Additional confusion may occur 
because the $+$ sign has been omitted, and transitions with rounded $A_1$/$A_2$ 
splitting of less than 0.1~kHz have been merged. These transitions are thus between 
two $A$ levels without parity designation. Moreover, since only 2 characters are 
available for each quantum number, $K$ values $\geq 10$ with negative parity are 
indicated by lower case characters; e.g. a2, a5, a8, and b1 are used for $-12$, $-15$, 
$-18$, and $-21$, respectively. The entry in HITRAN provides their common parity labels. 
It should be also mentioned that a process is under way to create a database version 
of the CDMS within the framework of the Virtual Atomic and Molecular Data Centre 
(VAMDC)\footnote{http://www.vamdc.eu/}~\cite{VAMDC_2010}. 
The official release of a test version is still pending because very many entries need 
to be imported properly, and in many cases necessary auxiliary files need to be created. 
One of the goals of VAMDC is providing easily interpretable quantum numbers.

Fairly recently, dipole moment and potential energy surfaces were reported, and selected 
ground state rotational energies have been provided~\cite{PH3_PES_2006}. The deviations 
between their energies and the present ones are fairly small. They drop at $J = 10$ from 
0.9164~cm$^{-1}$ for $K = 0$ to 0.4720~cm$^{-1}$ for $K = 9$. The deviations are larger 
at $J = 20$, where they drop from 3.4312~cm$^{-1}$ for $K = 0$ to 2.9680~cm$^{-1}$ for 
$K = 9$. Because of the extent of the experimental data, these deviations are probably 
predominantly due to insufficiencies in the calculated potential energy surface. 

More recently, a very extensive and very accurate line list has been created for 
PH$_3$~\cite{PH3_line_list_2013}. The authors found a remarkably good agreement between 
their transition frequencies and the ones in the first CDMS entry which had been created 
in October 2008 and which has been recalculated in February of 2013 by lowering the 
intensity cut-offs to provide weaker transition frequencies which in part access higher 
quantum numbers. These predictions correspond essentially to the ones derived from the 
alternative fit in the present work. 
The authors mentioned that the deviations are below 0.003~cm$^{-1}$ for the strongest 
100 transitions~\cite{PH3_line_list_2013}. As these transitions involve $J$ values 
between 3 and 14 and $K$ values from 0 to 9, which are all covered by experimental data, 
the deviations are probably again predominantly due to insufficiencies in the 
calculated potential energy surface. Moreover, differences in the predicted frequencies 
generated from the preferred fot or the alternative fit are very small for 
these transitions. Even more remarkable are the deviations at low quantum numbers. 
The $J_K = 1_0 - 0_0$ prediction from Ref.~\cite{PH3_line_list_2013} is only 2.8~MHz 
lower than the experimental frequency as well as the frequency calculated from 
either present fit. The deviations decrease first with increasing $J$, until 
the frequency of the $8_0 - 7_0$ transition is actually higher than the one predicted 
from the preferred fit. Similarly, the $2_{\pm1} - 2_{\mp2}$ transition frequency from 
Ref.~\cite{PH3_line_list_2013} is only 3.6~MHz lower than the one predicted from 
the preferred fit. Again, the deviations decrease first for increasing $J$ until 
the $10_{\pm1} - 10_{\mp2}$ transition frequency is higher in frequency. 
The agreement is still rather good if all transitions from the previous CDMS predictions 
are compared with those from Ref.~\cite{PH3_line_list_2013}; the authors give an rms value 
of 0.076~cm$^{-1}$. Because of the limitations in the experimental data as well as in the 
Hamiltonian model, it may well be that the predicted transition frequencies from the present 
study are eventually less reliable that those from Ref.~\cite{PH3_line_list_2013}. 

The intensities from Ref.~\cite{PH3_line_list_2013} agree very well with either prediction 
from the present fits for $\Delta K = 0$ transitions. The value for $\mu ^2$ from 
the former work is 3.9\,\% larger than the experimental value from Ref.~\cite{PH3_RF_1971}; 
this deviation is almost canceled by the neglect of the vibrational contributions to the 
partition function at 300~K in the present work. However, there are larger deviations 
between the calculated intensities of $\Delta K = 3$ transitions; e.g., the intensity 
of the $2_{\pm1} - 2_{\mp2}$ transition from Ref.~\cite{PH3_line_list_2013} is a factor 
of 1.85 weaker than the intensity from predictions generated from the preferred fit; 
it is even weaker by a factor of 2.5 if predictions generated from the alternative fit 
are used. Possible explanations for these deviations are insufficiencies in the calculations 
from Ref.~\cite{PH3_line_list_2013}, insufficiencies in the preferred Hamiltonian model, e.g. 
$\epsilon _K$ had to be constrained to zero, or the neglect of distortion corrections 
to the electric dipole moment. It may well be that each of these explanations is 
responsible in part.

%%%%%%%%%%%%%%%%%%%%%%%%%%%%%%%%%%%%%%%%%%%%%%%%%%%%%%%%%%%%%%%%%%%%%%%%%%%%%%%%%%%%%
%%%%%  Conclusion  %%%%%%%%%%%%%%%%%%%%%%%%%%%%%%%%%%%%%%%%%%%%%%%%%%%%%%%%%%%%%%%%%%
%%%%%%%%%%%%%%%%%%%%%%%%%%%%%%%%%%%%%%%%%%%%%%%%%%%%%%%%%%%%%%%%%%%%%%%%%%%%%%%%%%%%%

\section{Conclusion}
\label{Conclusion}

Improved spectroscopic parameters of PH$_3$ have been determined from recent very accurate 
rotational transition frequencies combined with previous data. Predictions for the 
ground state rotational spectrum of PH$_3$ generated from the preferred fit and employing 
the permanent dipole moment of 0.57395~D~\cite{PH3_RF_1971} will be provided in the catalog 
section of the CDMS\footnote{http://www.astro.uni-koeln.de/cdms/catalog}~\cite{CDMS_1,CDMS_2} 
and in the new edition of HITRAN\footnote{http://www.cfa.harvard.edu/HITRAN/}. 
Transition frequencies with assignments, uncertainties, and residuals between the observed 
frequencies and those calculated from the preferred parameter set as well as the predictions 
in the CDMS format will be provided as supplementary material. Line, parameter, and fit files 
along with various auxiliary files will be provided in the archive section of the CDMS catalog.
The old predictions together with available auxiliary files will be available in the archive section 
of the CDMS\footnote{http://www.astro.uni-koeln.de/site/vorhersagen/catalog/archive/PH3/version.1/}. 


%%%%%%%%%%%%%%%%%%%%%%%%%%%%%%%%%%%%%%%%%%%%%%%%%%%%%%%%%%%%%%%%%%%%%%%%%%%%%%%%%%%%%
%%%%%  acknowledgements  %%%%%%%%%%%%%%%%%%%%%%%%%%%%%%%%%%%%%%%%%%%%%%%%%%%%%%%%%%%%
%%%%%%%%%%%%%%%%%%%%%%%%%%%%%%%%%%%%%%%%%%%%%%%%%%%%%%%%%%%%%%%%%%%%%%%%%%%%%%%%%%%%%

\section*{Acknowledgements}

I thank Cristina Puzzarini for providing the data files associated 
with Ref.~\cite{PH3_rot_Lamb-Dip_2006}.
Moreover, I am very grateful to the Bundesministerium f\"ur Bildung 
und Forschung (BMBF) for financial support through project FKZ 50OF0901 
(ICC HIFI \textit{Herschel}). 

\appendix

\section*{Appendix A. Supporting information}

Supplementary data associated with this article can be found in the 
online version at http://dx.doi.org/10.1016/j.jqsrt.2013.xx.yyy.

%% References


%% \bibliographystyle{elsarticle-num}
%% \bibliography{<your-bib-database>}



%%%%%%%%%%%%%%%%%%%%%%%%%%%%%%%%%%%%%%%%%%%%%%%%%%%%%%%%%%%%%%%%%%%%%%%%%%%%%%%%%%%%%
%%%%%%%%%%%%%%%%%%%%%%%%%%%%%%%%%%%%%%%%%%%%%%%%%%%%%%%%%%%%%%%%%%%%%%%%%%%%%%%%%%%%%
%%%%%%%%%%%%%%%%%%%%%%%%%%%%%%%%%%%%%%%%%%%%%%%%%%%%%%%%%%%%%%%%%%%%%%%%%%%%%%%%%%%%%

\end{document}